\documentclass[pra,10pt,aps,superscriptaddress,twocolumn,amsmath,amssymb,nofootinbib,longbibliography]{revtex4-1}

\usepackage{bm}
\usepackage{natbib}
\usepackage[colorlinks]{hyperref}
\usepackage{xurl}
\usepackage{url}
\hypersetup{breaklinks=true,linkcolor=blue,citecolor=blue,filecolor=blue,urlcolor=blue}
\usepackage{orcidlink}

\usepackage{stackengine}
\stackMath

\usepackage{scalerel}
\newcommand*{\paral}{\stretchrel*{\parallel}{\perp}}
\usepackage{graphicx}
\usepackage{bm}
\usepackage{color}
\usepackage{soul}

\definecolor{green}{rgb}{0.19,0.64,0.54}
\definecolor{blue}{rgb}{0,0,1}
\definecolor{reddish}{rgb}{0.65, 0.2, 0.2}
\definecolor{darkgreen}{rgb}{0.2,0.7,0.3}
\definecolor{darkblue}{rgb}{0.3,0.40,0.48}
\definecolor{gray}{rgb}{.8,.8,.8}

\def\nat{Nature}

\begin{document}

\title{Light propagation in magnetoelectric materials: The role of optical coefficients in refractive index modulation}

\author{Vitorio A. \surname{De Lorenci}\,\orcidlink{0000-0001-5880-2207}}
\email{delorenci@unifei.edu.br}
\affiliation{Instituto de F\'{\i}sica e Qu\'{\i}mica, Universidade Federal de Itajub\'a, \\
Itajub\'a, Minas Gerais 37500-903, Brazil}
\author{Lucas T. \surname{de Paula}\,\orcidlink{0009-0002-1192-3532}}
\email{tobias.l@ufabc.edu.br}
\affiliation{Centro de Matemática, Computação e Cognição, Universidade Federal do ABC, \\
Santo Andr\'e, S\~ao Paulo 09210-170, Brazil}
\begin{abstract}
Investigations into optical phenomena associated with nonlinear magnetoelectric effects are attracting growing attention within the scientific community. Technologies constantly demand new materials capable of exhibiting precise and controllable responses to external electromagnetic fields. In this context, the optics of such materials is of remarkable importance. Here, working in a lossless and non-dispersive regime, electromagnetic wave propagation in materials presenting linear and nonlinear optical coefficients is investigated. We expand the discussion of the roles of nonlinear coefficients by examining special cases in which the contribution of the magnetoelectric optical coefficients $\alpha_{ij}$, $\beta_{ijk}$, and $\gamma_{ijk}$ to birefringence and nonreciprocal phenomena is elucidated. Notably, expressions that directly connect the magnetoelectric coefficients to the refractive indices of the medium are fully derived. These expressions enable the direct measurement of all components of each nonlinear magnetoelectric coefficient, providing an advancement over previous works. This development bridges theoretical models with experimental applications, offering possibilities for the optical characterization of magnetoelectric effects.
\end{abstract}

\maketitle

\section{Introduction}
Materials that magnetize through an applied electric field or polarize through an applied magnetic field are classified as magnetoelectric. They have been extensively studied in recent years and are now being used in the conception of new technological devices \cite{2008JAP...103c1101N,2015JPCM...27X4002O,2021APLM....9d1114L,2021APLM....9e0401H}. To give a few examples, controlling magnetization by means of an electric field at room temperature was reported to be significant for the design of energy-efficient spintronic devices \cite{2014Natur.516..370H}, and the use of magnetoelectric composites in flexible electronics was recently proposed to enable the development of technologies such as smart textiles, biosensors, and self-powered devices \cite{2023_Sasmal}. Also, studies exploring nonlinear magnetoelectric effects in layered structures have been reported, with possible applications in radio-frequency magnetometry \cite{PhysRevA.105.063509}. Some recent studies addressing magnetoelectricity have the objective, not only of measuring such effect \cite{chang}, but also of using the measurements to classify other types of physical features, such as skyrmions \cite{spaldin}. Ways to enhance the magnetoelectric phenomenon are also investigated \cite{mellado}.

The study of electromagnetic wave propagation in magnetoelectric systems has been a subject of prolonged consideration in the literature. Theoretical accounts of light propagation in these special materials were reported in the 1960s \cite{dell1962,fuchs1965,birss1967}, when the study of plane waves was considered in special regimes and with materials exhibiting specific symmetries. 
Wave propagation in bianisotropic media has also been a subject of investigation \cite{vytovtov2001bianisotropic,2016ITAP...64.5382L}. This topic is compelling in the context of uniaxial and chiral materials \cite{Lindell1994book,2010eabf.book.....M}, and also in metamaterials \cite{capolinobook}. We also mention the study of light propagation in a generic local and linear medium, with possible metamaterial realizations \cite{2016PhRvA..93a3844F}, the experimental observation of nonreciprocal transmission of a light ray in a magnetoelectric material \cite{PhysRevLett.123.077401}, and the analysis of reciprocal and nonreciprocal light propagation in the magnetoelectric antiferromagnet CuB${}_2$O${}_4$, which revealed a large variety of other interesting optical effects \cite{PhysRevB.104.184108}. Systems in lower dimensions have also been taken into consideration \cite{PhysRevA.106.063521,ying2022}.

Another intriguing facet of magnetoelectric materials lies in their nonlinear effects. Measurements of the second-order magnetoelectric effect were reported in the literature long ago~\cite{1969PMag...20.1087C,1971PSSBR..45..597C,MERCIER19771089}, and the topic is still of current interest. For instance, the emergence of second-order effects during field cooling in sputter-grown films was recently reported for a typical linear material~\cite{2021PhRvM...5i4406A}, and it was suggested that site-selective trace dopants are responsible for disrupting spacetime symmetries. These findings open new avenues for understanding symmetry-breaking mechanisms in magnetoelectric systems.
 
Nonlinear materials are characterized by their nonlinear response to external electromagnetic fields. In many practical situations, however, these external fields can be separated into two components: a strong, slowly varying (or static) field that induces polarization and magnetization phenomena in the optical material, and a weaker, rapidly varying wave field that propagates through it. When the latter is sufficiently small compared to the strong component, it satisfies a linear wave equation. In this work, we restrict our analysis to this regime, allowing the use of linear optics methods even though the background medium is nonlinear.

The primary goal of this study is to investigate the propagation of monochromatic plane waves in nonlinear magnetoelectric materials, considering optical coefficients that couple to electric and magnetic fields up to second order. A lossless and nondispersive regime is assumed. We adopt an approach in which the fields $\boldsymbol{E}$ and $\boldsymbol{B}$ are chosen to be thermodynamic variables for the polarization $\boldsymbol{P} $ and magnetization $\boldsymbol{M}$ vectors.
Previous analysis \cite{PhysRevA.105.023530} is generalized by the inclusion of all possible second-order nonlinear optical couplings, namely those related to polarization and magnetization phenomena activated by squared electric and magnetic fields, as well as the two possible cross-couplings. Furthermore, we also explore the mathematical convenience of using the magnetic field $\boldsymbol{B}$ as a variable for $\boldsymbol{P}$ and $\boldsymbol{M}$, instead of the auxiliary field $\boldsymbol{H}$.

In the next section, basic aspects of second-order, nonlinear materials are introduced.
The constitutive relations are given and subsequently used in Sec.~\ref{Plane-wave} to derive the eigenvalue problem associated with plane-wave propagation in a general second-order nonlinear magnetoelectric medium. Solutions to the eigenvalue problem are thus examined in Secs.~\ref{caseb} and \ref{casec}. Expressions relating the nonlinear magnetoelectric coefficients to possible observable quantities are also presented. A special case where both linear and nonlinear effects modulate the refractive index of the material is examined in Sec.~\ref{mixingeffects}. Next, in Sec.~\ref{linearmedium} the same ideas are applied to the case of linear magnetoelectric materials. Particularly, Sec~\ref{linearmedium} revisits a solution discussed long ago \cite{fuchs1965}, in which a treatment based on $\boldsymbol{E}$ and $\boldsymbol{H}$ fields was implemented. The results suggest that the magnetoelectric effect has a simpler formulation in the formalism based on $\boldsymbol{E}$ and $\boldsymbol{B}$ fields. 
Final remarks and conclusions are presented in Sec.~\ref{final}.
For comparison purposes, the treatment of the linear case using the auxiliary field $\boldsymbol{H}$ \cite{fuchs1965} in the expansion of the polarization and magnetization is presented briefly in a self-contained manner in Appendix~\ref{appendix}. A comparative analysis of different definitions of magnetic susceptibility is presented in Appendix~\ref{magsus}. It is argued that a definition based on magnetization $\boldsymbol{M}(\boldsymbol{B})$, instead of $\boldsymbol{M}(\boldsymbol{H})$, leads to more natural expressions for certain results in electromagnetism.
Finally, a preliminary discussion of the applicability of the index-ellipsoid method in the context of nonlinear magnetoelectric materials is presented in Appendix~\ref{ellipsoid}.

Following the Einstein convention for sums, repeated indices in a monomial indicate summation. The three-dimensional Levi-Civita symbol $\epsilon_{ijk}$ is a completely antisymmetric rank-3 quantity, defined by $\epsilon_{123}=1$. Thus, the curl of an arbitrary vector $\boldsymbol{V}$ is written in component notation as $\epsilon_{ijk}\partial_j V_k$, where $\partial_j$ means the partial derivative with respect to coordinate $x_j$. The divergence of $\boldsymbol{V}$ is simply $\partial_i V_i$. Additionally, $\partial_t$ will be used to denote a time derivative. The Kronecker delta, denoted as $\delta_{ij}$, takes a value of 1 when $i=j$ and 0 otherwise.
The wave vector is denoted by $\boldsymbol{ q}$, with Cartesian components $q_i$. This vector can also be expressed in terms of its dimensionless directional unit vector $\boldsymbol{\kappa}$ as ${\boldsymbol q} = q \, \boldsymbol{\kappa}$, such that $q_i = q \kappa_i$, $q_i q_i = q^2$, and $\kappa_i \kappa_i = 1$.

\section{The constitutive relations}
\label{FEB}
The optical properties of a medium are mainly associated with the way it can be polarized and magnetized by applied electromagnetic fields. The effect of these fields over a nonconducting medium is basically to rearrange charge and magnetic moment distributions. Energy will be stored in the medium as a consequence of the presence of these fields. The analysis here is restricted to the realm of solid crystalline materials, such that the volume $V$ and temperature $T$ of the system are constants, the latter being externally controlled. Thus, the free energy is the function that is minimized in thermodynamic processes that can occur in such systems. The polarization $\boldsymbol{P}$ and magnetization $\boldsymbol{M}$ vectors can be obtained by means of Maxwell's relations involving the free-energy density $F$ of the material. The electric $\boldsymbol{ E}$ and magnetic $\boldsymbol{ B}$ fields or the auxiliary $\boldsymbol{D}$ and $\boldsymbol{H}$ fields or even a mix of them, as is usually described in the literature, can be chosen as thermodynamic variables for $F$.
Some consequences of using the fundamental or auxiliary fields in the expansion of the free-energy density are discussed in \cite{2009EPJB...71..299R} (see also the references therein).

For a solid crystalline material at a given temperature $T$, the free-energy density can be expanded in terms of electric and magnetic fields as \cite{rivera1994a,2009EPJB...71..299R}
\begin{align}
F\left(\boldsymbol{E}, \boldsymbol{B};T\right) =
\; F_{0}-P_i^{\scriptscriptstyle {S}} E_{i}-M_i^{\scriptscriptstyle {S}} B_{i} -\tfrac{1}{2}\varepsilon_0\chi_{i j}^{\scriptscriptstyle \boldsymbol{E\!B}} E_{i} E_{j}
\nonumber \\
 -\tfrac{1}{2\mu_0}\tilde\chi_{i j}^{\scriptscriptstyle \boldsymbol{E\!B}} B_{i} B_{j} -\alpha_{i j}^{\scriptscriptstyle \boldsymbol{E\!B}} E_{i} B_{j}-\tfrac{1}{3!}\varepsilon_0 \chi_{ijk}^{\scriptscriptstyle \boldsymbol{E\!B}}E_i E_j E_k
\nonumber\\
 -\tfrac{1}{3!}\tilde\chi_{i j k}^{\scriptscriptstyle \boldsymbol{E\!B}}B_i B_j B_k-\tfrac{1}{2}\beta_{ijk}^{\scriptscriptstyle \boldsymbol{E\!B}}E_i B_j B_k-\tfrac{1}{2}\gamma_{ijk}^{\scriptscriptstyle \boldsymbol{E\!B}}B_i E_j E_k.
\nonumber
\end{align}
The above Taylor expansion was truncated in order to consider only optical effects up to second order. The free-energy density of the material in the absence of external fields is $F_0 \doteq F(0,0;T)$. 
The coefficients $P_i^{\scriptscriptstyle {S}}$ and $M_i^{\scriptscriptstyle {S}}$ represent the $i$th components of the spontaneous polarization and magnetization vectors, 
respectively. The dimensionless rank-2 coefficients $\chi_{i j}^{\scriptscriptstyle \boldsymbol{E\!B}}$ and $\tilde\chi_{i j}^{\scriptscriptstyle \boldsymbol{E\!B}}$ denote the first-order electric and magnetic susceptibilities of the medium, respectively, while the rank-3 coefficients $\chi_{ijk}^{\scriptscriptstyle \boldsymbol{E\!B}}$ and $\tilde\chi_{i j k}^{\scriptscriptstyle \boldsymbol{E\!B}}$ represent the second-order electric and magnetic susceptibilities, 
respectively. In the magnetoelectric sector, $\alpha_{i j}^{\scriptscriptstyle \boldsymbol{E\!B}}$ represents the linear magnetoelectric coefficient, 
while $\beta_{ijk}^{\scriptscriptstyle \boldsymbol{E\!B}}$ and $\gamma_{ijk}^{\scriptscriptstyle \boldsymbol{E\!B}}$ account for the second-order magnetoelectric effects. It should be noted that the dimensions of $\mu_0 \tilde\chi_{i j k}^{\scriptscriptstyle \boldsymbol{E\!B}}$, $\mu_0 \alpha_{i j}^{\scriptscriptstyle \boldsymbol{E\!B}}$, $\mu_0^2\beta_{ijk}^{\scriptscriptstyle \boldsymbol{E\!B}}$ and $\mu_0\gamma_{ijk}^{\scriptscriptstyle \boldsymbol{E\!B}}$ are, respectively,  ${\rm m}~ {\rm A}^{-1}$, ${\rm s} ~{\rm m}^{-1}$, ${\rm s}~ {\rm A}^{-1}$ and ${\rm s}~ {\rm V}^{-1}$.
Note that, depending on the symmetry of the system, not all coefficients in the above expansion will be present. For example, under a space-reversal transformation $(\vec r \to -\vec r)$, odd powers of the electric field will change sign. Therefore, if the system is symmetric under such a transformation, the coefficients of all those terms will be identically zero.

It should be emphasized at this point that the optical coefficients that couple to the magnetic field in the expansion of $F\left(\boldsymbol{E}, \boldsymbol{B};T\right)$ should not be mistaken for those that appear in a description based on $F\left(\boldsymbol{E}, \boldsymbol{H};T\right)$. For instance, the magnetic susceptibility is traditionally defined as the coefficient of the auxiliary field $\boldsymbol{H}$ in the expression for magnetization $\boldsymbol{M}$, which is a consequence of the latter description.
Similarly, the linear magnetoelectric coefficient defined above should not be confused with the one that most often appears in the literature, which is denoted by $\alpha_{i j}^{\scriptscriptstyle \boldsymbol{E\!H}}$ the Appendix \ref{appendix}. The same distinctions hold for all the nonlinear coefficients, except for $\chi_{ijk}$, as it couples only with the electric field.

Hereafter, in order to maintain simpler notation, the superscript ${\boldsymbol{E\!B}}$ related to the optical coefficients appearing in the expansion of $F(\boldsymbol{E},\boldsymbol{B};T)$ will be omitted. 
The polarization and magnetization vectors can be directly obtained by means of the thermodynamic relations
\begin{subequations}
\label{p&m}
\begin{align}
\label{p&}
P_{i}=&-\frac{\partial F}{\partial E_{i}}=P_i^{\scriptscriptstyle {S}}+\varepsilon_{0} \chi_{i j} E_{j}+\alpha_{i j} B_{j}\\  
&+\tfrac{1}{2}\varepsilon_0\chi_{ijk}E_j E_k+\tfrac{1}{2}\beta_{ijk}B_jB_k+\gamma_{jik}B_j E_k,
\nonumber\\
\label{&m}
M_{i}=&-\frac{\partial F}{\partial B_{i}}=M_i^{\scriptscriptstyle {S}}+ \frac{\tilde\chi_{i j} B_{j}}{\mu_{0}}+\alpha_{j i} E_{j}\\
&+\tfrac{1}{2}\tilde\chi_{ijk}B_j B_k+\beta_{jik}E_j B_k+\tfrac{1}{2}\gamma_{ijk}E_j E_k.
\nonumber
\end{align}
\end{subequations}

The auxiliary fields $D_{i}$ and $H_{i}$ are given by $D_{i}  =\varepsilon_{0} E_{i}+P_{i}$ and $H_{i}=(B_{i}/\mu_0) - M_{i}$. Thus, the time-domain constitutive relations that follow from this formalism are given by
\begin{subequations}
\label{dh1} 
\begin{align}
\label{d1}
D_{i} =\,& P_i^{\scriptscriptstyle {S}} + \varepsilon_{i j} E_{j}+\tfrac{1}{2}\varepsilon_0\chi_{ijk}E_kE_j
\\
&+\left(\alpha_{i j}+\tfrac{1}{2}\beta_{ijk}B_k+\gamma_{jik}E_k\right)B_j,
\nonumber
\\
\label{h1}
H_{i}=\,&M_i^{\scriptscriptstyle {S}}+\bar\mu_{i j} B_{j}-\tfrac{1}{2}\tilde\chi_{ijk}B_jB_k
\\
&-\left(\alpha_{j i}+\beta_{jik}B_k+\tfrac{1}{2}\gamma_{ijk}E_k\right)E_j,
\nonumber
\end{align}
\end{subequations}
where the electric permittivity $\varepsilon_{i j}$ and the inverse magnetic permeability $\bar\mu_{i j}$ tensors are defined as
\begin{subequations} \label{epsilonmu} 
\begin{align}
\varepsilon_{i j}&=\varepsilon_{0}\left(\delta_{i j}+\chi_{i j}\right), 
\\
\bar\mu_{i j}&=\tfrac{1}{\mu_{0}}\left(\delta_{i j}-\tilde\chi_{i j}\right), \label{varmu}
\end{align} 
\end{subequations}
and the magnetic permeability $\mu_{i j}$ is defined such that $\bar\mu_{i j}\mu_{j k} = \delta_{ik}$. 

\section{Plane waves in a nonlinear magnetoelectric medium}
\label{Plane-wave}
In the absence of free sources of charge and current densities, Maxwell's equations are given by the two null divergences, $\partial_{i}D_{i}=0$ and $\partial_{i}B_{i}=0$, and the curl equations $\epsilon_{ijk}\partial_{j}E_{k}=-\partial_{t}B_{i}$  and $\epsilon_{ijk}\partial_{j}H_{k}=\partial_{t}D_{i} $. The total fields in these equations are decomposed into the sum of a strong and nearly constant background part, which induces polarization and magnetization in the dielectrics, and a weak and rapidly varying part, composed of the wave fields $E^{\omega}_j = e_j \exp(i\phi)$ and $B^{\omega}_j = b_j \exp(i\phi)$, where $\phi = q_j x_j -\omega t$ is the phase, with $q_j$ being the wave vector and $\omega$ being the angular frequency of the plane-wave solutions, and $e_j$ and $b_j$ are the wave-polarization vectors. Thus, $\partial E_i \approx \partial E^{\omega}_i$. Using this prescription in the two curl equations, together with the constitutive relations given by Eq.~(\ref{dh1}), and eliminating $b_j$ in favor of $e_j$, namely, $b_{i}=\tfrac{1}{\omega}\epsilon_{i j k}\, q_{j}\, e_{k}$, we can straightforwardly obtain the eigenvalue equation $Z_{ij}e_j = 0$, with the Fresnel tensor $Z_{ij}$ being defined by
\begin{align}
Z_{ij} =\; & \theta_{i j} v^{2} -2  \epsilon_{ln(i}\zeta_{j)n}\kappa_l v -\epsilon_{iln}\epsilon_{jrs}\lambda_{ns}\kappa_{r}\kappa_{l},
\label{zijnl}
\end{align}
where $v$ is the magnitude of the phase velocity ($\boldsymbol{v} = v \boldsymbol{\kappa}$) of the plane wave, defined by $v = \omega/q$, with $q=\sqrt{q_i q_i}$, and
\begin{subequations}
\begin{align}
    &\theta_{ij}=\varepsilon_{ij}+\varepsilon_0\chi_{ijk}E_k+\gamma_{kij}B_k,\\
    &\zeta_{ij}=\alpha_{ij}+\beta_{ijk}B_k+\gamma_{jik}E_k,\label{zetacoe}\\
    &\lambda_{ij}=\bar\mu_{ij}-\tilde\chi_{ijk}B_k-\beta_{kij}E_k.
\end{align}
\label{thzela}
\end{subequations}
Note that $\zeta_{ij}$ is a purely magnetoelectric tensor, while $\theta_{ij}$ and $\lambda_{ij}$ mix electric, magnetic, and magnetoelectric effects. However, when no external electric or magnetic fields are present, the magnetoelectric effect is solely represented by $\zeta_{ij}$.

General solutions for wave propagation in such a magnetoelectric medium can be obtained by solving the eigenvalue problem stated above. Particularly, the phase velocities are the solutions one obtains by equating the eigenvalues of $Z_{ij}$ to zero or, perhaps more directly, by solving $\det |Z_{ij}| =0$ for $v$ \cite{PhysRevA.105.023530}. Furthermore, the polarization vectors of the plane waves are given by the kernel of $Z_{ij}$; i.e., they are the eigenvectors of $Z_{ij}$ with null eigenvalues.

An alternative approach to describing light propagation in material media involves the use of the index ellipsoid \cite{landau1984electrodynamics,1999poet.book.....B}. However, incorporating magnetoelectric effects into this framework presents significant challenges, primarily due to the lack of symmetry in the optical coefficients, which complicates the geometric interpretation usually associated with the index ellipsoid. This issue is briefly discussed in Appendix \ref{ellipsoid}.

\subsection{Case with \texorpdfstring{$\beta_{ijk}$}{beta\_ijk}}
\label{casebeta}

\label{caseb}
In the remainder of this section, the analysis is restricted to materials whose linear sector is isotropic, namely,  $\varepsilon_{ij}=\varepsilon \delta_{ij}$, $\bar\mu _{ij}=\mu\delta_{ij}$, and $\alpha_{ij}=\alpha \delta_{ij}$. 

Let us consider a material in which the only significant nonlinear contribution is given by the coefficients $\beta_{ijk}$. In addition, we study the scenario in which the magnetic field is turned off, such that the $\zeta_{ij}$ coefficients do not contribute to the material response. The solutions for the phase velocity in an arbitrary direction, up to the first-order contributions in the nonlinear coefficients, are given by
\begin{eqnarray}
    &(v_{\pm})^2=\frac{1}{\varepsilon\mu}\Big[1-\frac{\mu}{4}\beta_{ij}I_{ij}&
    \nonumber\\
    &\pm\frac{\mu}{4}\sqrt{(\beta_{ij}I_{ij})^2+2\left(\beta_{ik}\beta_{jk}-\beta_{ij}\beta_{kk}\right)\left(I_{ij}-\kappa_{i}\kappa_{j}\right)}\Big]^2,&\nonumber\\
    \label{vb}
\end{eqnarray}
where we have defined $\beta_{ij}=\beta_{kij}E_k$ and $I_{ij}=\delta_{ij}-\kappa_{i}\kappa_{j}$.
Note that the magnetoelectric effect acts effectively like the magnetic Pockels effect but is induced by an applied electric field. This effect artificially breaks the isotropy of the medium and activates birefringence. 

The refractive index, namely, $n=c/v$, in the three main directions of propagation can be obtained directly from Eq.~(\ref{vb}) and results, up to first order in $\beta_{ijk}$, in
\begin{align}
\resizebox{\columnwidth}{!}{$
n^x_{\pm}=n_0\left[1+\frac{\mu}{4}\left(\beta_{22}+\beta_{33}\pm\sqrt{(\beta_{22}-\beta_{33})^2+4(\beta_{23})^2}\right)\right],
 $} \nonumber \\  
\resizebox{\columnwidth}{!}{$
n^y_{\pm}=n_0\left[1+\frac{\mu}{4}\left(\beta_{11}+\beta_{33}\pm\sqrt{(\beta_{11}-\beta_{33})^2+4(\beta_{13})^2}\right)\right],
 $} \nonumber \\ 
\resizebox{\columnwidth}{!}{$
n^z_{\pm}=n_0\left[1+\frac{\mu}{4}\left(\beta_{11}+\beta_{22}\pm\sqrt{(\beta_{11}-\beta_{22})^2+4(\beta_{12})^2}\right)\right],
 $} \nonumber  
\end{align}
where we have defined the ordinary refractive index of the medium, $n_0\doteq c\sqrt{\varepsilon\mu}$, which characterizes the propagation of light in the absence of external fields. 

We now construct the quantity
\begin{equation}
\Delta^d=\frac{(n_+^d-n_0)+(n_-^d-n_0)}{n_0},
\label{delta}
\end{equation}
where the superscript $d=x, y, z$ indicates the spatial direction. For a given direction, this quantity measures the average deviation of $n_+^d$ and $n_-^d$ from $n_0$, normalized by $n_0$.

With the above results it is straightforward to obtain the following nonlinear magnetoelectric coefficients
\begin{eqnarray}
&\beta_{11}&=\beta_{k11}E_k=\tfrac{1}{\mu}(-\Delta^x+\Delta^y +\Delta^z),
\nonumber\\
&\beta_{22}&=\beta_{k22}E_k=\tfrac{1}{\mu}(\Delta^x-\Delta^y +\Delta^z),
\nonumber\\
&\beta_{33}&=\beta_{k33}E_k=\tfrac{1}{\mu}(\Delta^x +\Delta^y -\Delta^z).
\nonumber
\end{eqnarray}
Setting the direction of the electric field allows us to compute the full diagonal sector of the last two indices of $\beta_{ijk}$, which consists of nine components. 

Now we define the birefringence coefficient
\begin{equation}
    \Theta^d=\left|\frac{n_+^d-n_-^d}{n_0}\right|,
    \label{theta}
\end{equation}
which measures the relative difference of the refractive indices with respect to $n_0$.

For the three main directions we find
\begin{eqnarray}
&(\beta_{12})^2=(\beta_{k12}E_k)^2=\frac{(\Theta^z)^2-(\Delta^x -\Delta^y)^2}{\mu^2},&
\nonumber\\
&(\beta_{13})^2=(\beta_{k13}E_k)^2=\frac{(\Theta^y)^2-(\Delta^x -\Delta^z)^2}{\mu^2},&
\nonumber\\
&(\beta_{23})^2=(\beta_{k23}E_k)^2=\frac{(\Theta^x)^2-(\Delta^z -\Delta^y)^2}{\mu^2}.&
\nonumber
\end{eqnarray}
Note that the above equations allow for the determination of all remaining components of $\beta_{ijk}$.

There are numerous experimental results reported in the literature (see, for instance, Refs.~\cite{1965PhRv..138.1218H,1985Tabares,doi:10.1080/00150199408213358}) concerning this second-order magnetoelectric contribution. As an example, if we take the reported values for $\beta_{ijk}$ in
${\rm Cr}_3{\rm B}_7\rm{O}_{13}{\rm Cl}$ ($\beta^{\scriptscriptstyle \boldsymbol{E\!H}}_{322} = 4\times10^{-19} {\rm s\,~A}^{-1}$ and $\beta^{\scriptscriptstyle \boldsymbol{E\!H}}_{333} = 1.5\times10^{-18} {\rm s\,~A}^{-1}$ at 4.2K \cite{doi:10.1080/00150199408213358}) as a reference, we find that, in the absence of an external magnetic field, the magnitude of the birefringence effect will be given by
\begin{equation}
\Theta^x \approx 10^{-7} \frac{\mu}{\mu_0}\left(\frac{E_k}{10^5{\rm V}{\rm m}^{-1}}\right)\frac{\sqrt{(\beta_{k22}-\beta_{k33})^2+4(\beta_{k23})^2}}{10^{-6}{\rm A m^2 s^{-1} V^{-2}}},
\nonumber
\end{equation}
which coincides with the estimates obtained earlier when a description based on $F\left(\boldsymbol{E}, \boldsymbol{H};T\right)$ was implemented~\cite{PhysRevA.105.023530}, as expected.

The birefringence coefficient defined in Eq.~(\ref{theta}) can be generalized for an arbitrary direction by calculating the refractive indices with the complete solutions for the phase velocities described in Eq.~(\ref{vb}). The resulting expression is a function of the spherical angles $\theta$ and $\varphi$ as 
\begin{align}
\Theta(\theta,\varphi)=\left|\frac{n_+(\theta,\varphi)-n_-(\theta,\varphi)}{n_0}\right|.
\label{Thg}
\end{align}
\begin{figure}[b]
\includegraphics[width=\linewidth]{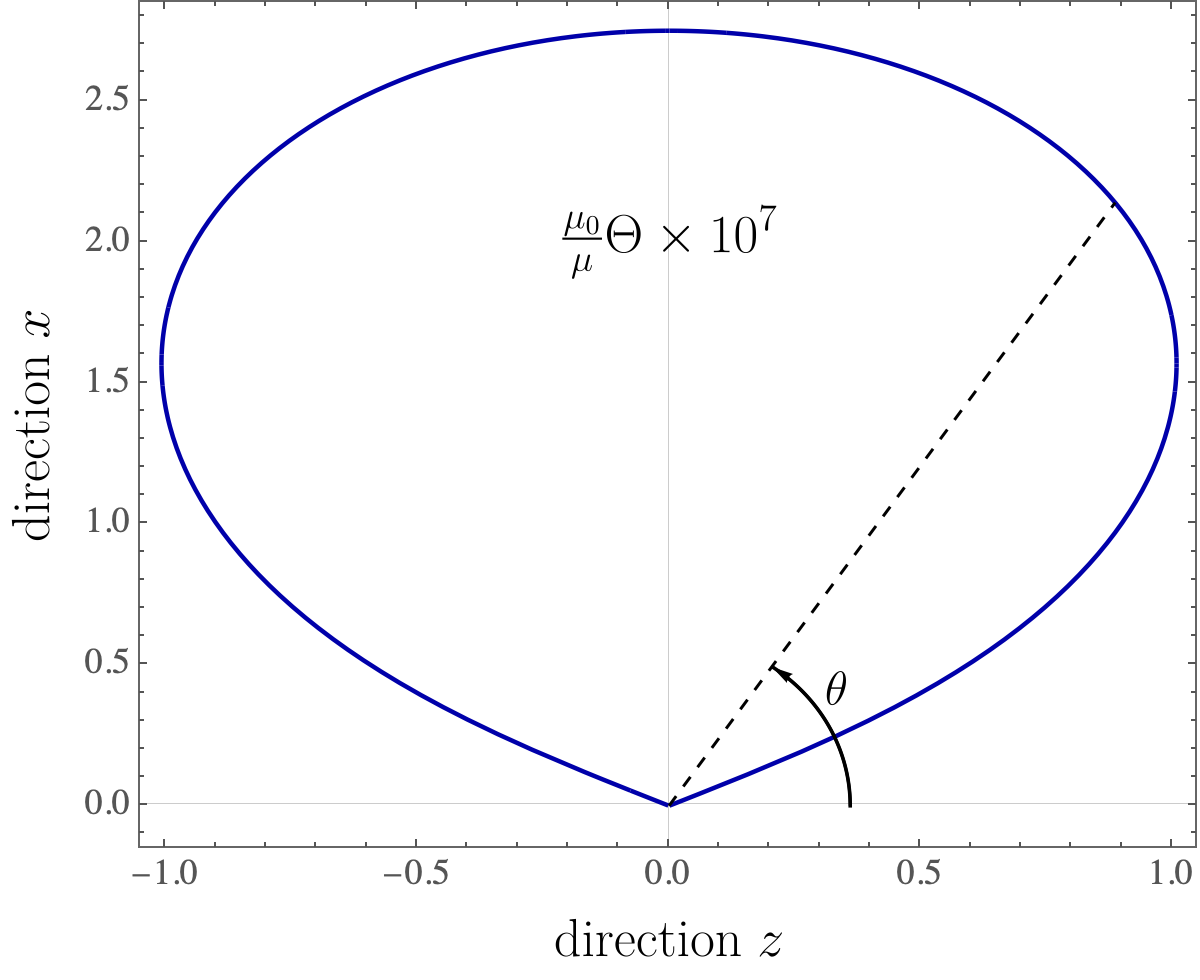}
\caption{The birefringence coefficient $\Theta(\theta)$ is shown for arbitrary directions (specified by the spherical angle $\theta$ ) in the $xz$ plane. Here, only $\beta_{322}$ and $\beta_{333}$ are assumed to be nonzero. This solution is expected to describe the contribution to the birefringence phenomenon associated with the $\beta_{ijk}$ optical coefficients in a ${\rm Cr}_3{\rm B}_7\rm{O}_{13}{\rm Cl}$ system \cite{doi:10.1080/00150199408213358}.
}
\label{figure1}
\end{figure}
In order to exhibit one case, in Fig.~\ref{figure1} we examine the particular system where the propagation is constrained to the $xz$ plane and $\beta_{322}$ and $\beta_{333}$ are assumed to be the only significant magnetoelectric contributions, with an applied electric field of $4.5\times 10^5$ V/m \cite{doi:10.1080/00150199408213358}. As the Fig.~\ref{figure1} shows, the birefringence effect occurs only within an angular opening of about $\pi/5$ rad around $\theta = \pi/2$, and it achieves its maximum at $\theta = \pi/2$, where $\Theta(\pi/2) \approx 2.7\times 10^{-7}(\mu/\mu_0)$. Any other configuration can be implemented by following the same procedure.

The above results should be compared with those presented in a previous publication \cite{PhysRevA.105.023530}, where the auxiliary field $\boldsymbol H$ was used in the expansions of the polarization and magnetization. Following the notation used here, the coefficients $\beta_{ijk}$ in that paper would be identified as $\beta_{ijk}^{\scriptscriptstyle \boldsymbol{E\!H}}$, whose physical dimension is $\rm s~ \rm A^{-1}$. In the particular case examined here, where the external magnetic field is absent and only this type of second-order magnetoelectric contribution is kept, we can identify $\beta_{ijk}=\tfrac{1}{\mu^2}\beta_{ijk}^{\scriptscriptstyle \boldsymbol{E\!H}}$.
Direct inspection shows that the solutions generally do not coincide. The origin of the difference is a little subtle. 
In the approach based on $\boldsymbol M(\boldsymbol H)$  \cite{PhysRevA.105.023530}, the constitutive relations were obtained after an expansion of the auxiliary field $\boldsymbol H$ in terms of $\boldsymbol B$ was implemented, keeping only first-order contributions in $\beta_{ijk}$. Thus, the eigenvalue equation turned out to be a linear function of the magnetoelectric coefficients. However, higher-order terms in the eigenvalue equation could give rise to first-order terms in the phase-velocity solutions. In contrast, in our approach, the assumption 
$\boldsymbol M=\boldsymbol M(\boldsymbol B)$ leads to a Fresnel tensor, as given by Eq.~(\ref{zijnl}), which is an exact and linear expression involving all magnetoelectric coefficients. The approximation used in the previous approach resulted in the emergence of an ordinary velocity, which, however, does not correspond to a propagating mode, except in some special configurations, for instance, in the absence of external fields, or when the second term inside the square root in Eq.~(\ref{vb}) vanishes or at least is negligible.

\subsection{Case with \texorpdfstring{$\gamma_{ijk}$}{gamma\_ijk}}

\label{casec}
Now, let us assume that the only significant nonlinear coupling is given by $\gamma_{ijk}$ and also that the external electric field is turned off. In this scenario, the solutions for the phase velocities, up to first order terms in $\gamma_{ijk}$, are given by
\begin{eqnarray}
   & (v_{\pm})^2=\frac{1}{\varepsilon\mu}\Big[1-\frac{1}{4\varepsilon}\gamma_{ij}I_{ij}& 
   \nonumber\\
   &\pm\frac{1}{4\varepsilon}\sqrt{(\gamma_{ij}I_{ij})^2+2\left(\gamma_{ik}\gamma_{jk}-\gamma_{ij}\gamma_{kk}\right)\left(I_{ij}-\kappa_{i}\kappa_{j}\right)}\Big]^2,&
   \nonumber\\ \label{vg}
\end{eqnarray}
where we defined $\gamma_{ij}\doteq\gamma_{kij}B_k$.

It can be inferred from these solutions that the nonlinear coupling of $\gamma_{ijk}$ with the magnetic field induces a breaking of the isotropy of the electric permittivity, leading to the emergence of an artificial birefringence phenomenon. This is analogous to the electric Pockels effect, except that now it is induced by the presence of an applied magnetic field.

The refractive indices for the three main directions are given by
\begin{align}
\resizebox{\columnwidth}{!}{$
n^x_{\pm}=n_0\left[1+\frac{1}{4\varepsilon}\left(\gamma_{22}+\gamma_{33}\pm\sqrt{(\gamma_{22}-\gamma_{33})^2+4(\gamma_{23})^2}\right)\right],
$} \nonumber \\
\resizebox{\columnwidth}{!}{$
n^y_{\pm}=n_0\left[1+\frac{1}{4\varepsilon}\left(\gamma_{11}+\gamma_{33}\pm\sqrt{(\gamma_{11}-\gamma_{33})^2+4(\gamma_{13})^2}\right)\right],
 $}\nonumber \\ 
\resizebox{\columnwidth}{!}{$
n^z_{\pm}=n_0\left[1+\frac{1}{4\varepsilon}\left(\gamma_{11}+\gamma_{22}\pm\sqrt{(\gamma_{11}-\gamma_{22})^2+4(\gamma_{12})^2}\right)\right].
 $} \nonumber 
\end{align}

Now, like for the case with $\beta_{ijk}$, using Eq.~(\ref{delta}), we find 
\begin{eqnarray}
&\gamma_{11}&=\gamma_{k11}B_k=\varepsilon(-\Delta^x+\Delta^y +\Delta^z),
\nonumber\\
&\gamma_{22}&=\gamma_{k22}B_k=\varepsilon(\Delta^x-\Delta^y +\Delta^z),
\nonumber\\
&\gamma_{33}&=\gamma_{k33}B_k=\varepsilon(\Delta^x +\Delta^y -\Delta^z),
\nonumber
\end{eqnarray}
and using Eq.~(\ref{theta}), the remaining coefficients can be obtained,
\begin{eqnarray}
&(\gamma_{23})^2=(\gamma_{k23}B_k)^2=\varepsilon^2\left[(\Theta^x)^2-(\Delta^z -\Delta^y)^2\right],&
\nonumber\\
&(\gamma_{13})^2=(\gamma_{k13}B_k)^2=\varepsilon^2\left[(\Theta^y)^2-(\Delta^x -\Delta^z)^2\right],&
\nonumber\\
&(\gamma_{12})^2=(\gamma_{k12}B_k)^2=\varepsilon^2\left[(\Theta^z)^2-(\Delta^x -\Delta^y)^2\right].&
\nonumber
\end{eqnarray}

The first measurements of these coefficients were reported in the late 1960, based on yttrium iron garnet systems \cite{1967PMag...16..487O,1969PMag...20.1087C,1971PSSBR..45..597C}, where values of $\mu_0 c \gamma_{311}$ of the order of $10^{-11} {\rm m}~{\rm V}^{-1}$ were found. Using this value as a reference, the birefringence effect produced by this type of second-order magnetoelectric contribution can be estimated by
\begin{equation}
\Theta^x \approx 5.65\times 10^{-3} \frac{\varepsilon}{\varepsilon_0}\left(\frac{B_k}{1{\rm T}}\right)\frac{\sqrt{(\gamma_{k22}-\gamma_{k33})^2+4(\gamma_{k23})^2}}{10^{-13}{\rm m A V^{-2}}}.
\nonumber
\end{equation}
Estimates of the effect in the other orthogonal directions can be obtained by simple permutation of the subscripts.
\begin{figure}[b]
\includegraphics[width=\linewidth]{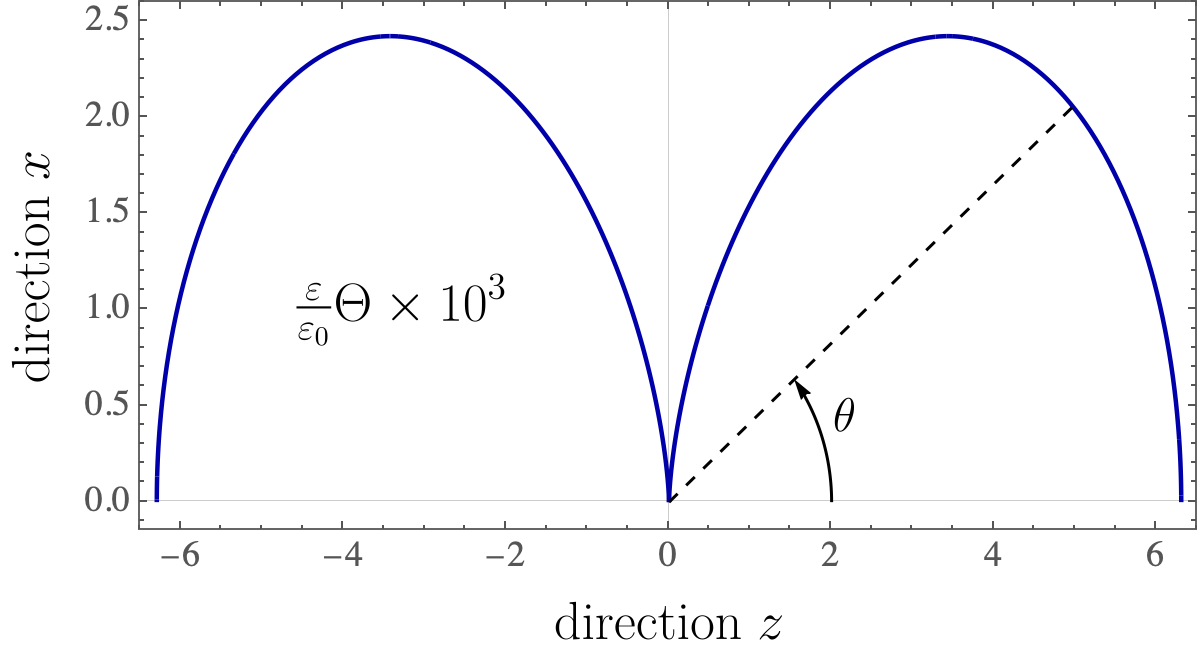}
\caption{The birefringence coefficient $\Theta(\theta)$ is shown for arbitrary directions in the $xz$ plane ($x>0$). Here, the only significant magnetoelectric contributions are assumed to be $\mu_0\gamma_{k11} \approx 7\times 10^{-20} {\rm s}~{\rm V}^{-1}$, which correspond to the values reported in the yttrium iron garnet system~\cite{1969PMag...20.1087C}, with an applied magnetic field of 1~T. }
\label{figure2}
\end{figure}

As discussed in Sec.~\ref{casebeta}, the birefringence coefficient can be studied in arbitrary directions by plugging the refractive indices from Eq.~(\ref{vg}) into $\Theta(\theta,\varphi)$, defined by Eq.~(\ref{Thg}).
A representative configuration is explored in Fig.~\ref{figure2} for a system where only $\gamma_{k11}$ ($k=1,2,3$) is significant. The wave propagation is set in the $xz$ plane, and a magnetic field of 1~T is present. As Fig.~\ref{figure2} reveals, the birefringence effect achieves its maximum about $\theta = 0$ and $\theta = \pi$, where $\Theta(0) \approx 6.3\times 10^{-3}(\varepsilon_0/\varepsilon)$. Note that there is no birefringence at $\theta = \pi / 2$.

In the previously examined models, the linear magnetoelectric coefficient was assumed to be isotropic. Direct inspection of Eq.~(\ref{zijnl}) shows that, although $\zeta_{ij} = \alpha \delta_{ij}$ is nonzero, it does not contribute to the Fresnel tensor and thus does not play a role in any observable effects. Only nonlinear magnetoelectric effects are involved in the results. However, in a scenario where $\beta_{ijk}$ is the only significant nonlinear effect and the electric field is turned off, $\zeta_{ij}$ does contribute to the effects. In this case, it is noteworthy that if we set the magnetic field in the \textit{a}th direction, say, $B_i = B \delta_{ia}$, and select the system such that $\beta_{ija} = \operatorname{diag}(\beta_{11a}, \beta_{22a}, \beta_{33a})$, the wave propagation description becomes fully equivalent to the linear magnetoelectric case with a diagonal coefficient $\alpha_{ij} = \operatorname{diag}(\alpha_{1}, \alpha_{2}, \alpha_{3})$. Similarly, when $\gamma_{ijk}$ is nonzero and the magnetic field is turned off, the same reasoning applies. The issue of combining linear and nonlinear effects will be examined in the next section. 

\section{Mixing linear and nonlinear magnetoelectric effects}
\label{mixingeffects}
When $\zeta_{ij}$ contains nonlinear contributions, finding solutions for the wave equation becomes a difficult task. However, there are some configurations for which exact solutions can be found. That can be achieved by conveniently adjusting the external fields and assuming a material medium that exhibits certain symmetries. 

First, let us assume that the physical system possesses a particular symmetry that allows us to align the principal axes of $\varepsilon_{i j}$, $\bar\mu_{ i j}$, and $\alpha_{ij}$. 
Next, the three Cartesian axes are chosen to coincide with these principal axes. In such a case these optical coefficients are diagonal tensors, which hereafter will be represented by 
\begin{subequations}
\label{coefs}
\begin{align}
    \varepsilon_{i j}&=\operatorname{diag}\left(\varepsilon_{1}, \varepsilon_{2}, \varepsilon_{3}\right),
    \label{varepdiag} 
    \\
    \bar\mu_{i j}&=\operatorname{diag}\left({\tfrac{1}{\mu_{1}}}, {\tfrac{1}{\mu_{2}}},{\tfrac{1}{\mu_{3}}}\right),
    \label{mbardiag}
    \\
    \alpha_{i j}&=\operatorname{diag}\left(\alpha_{1}, \alpha_{2}, \alpha_{3}\right).
    \label{alphadiag}
\end{align}
\end{subequations}
Additionally, we restrict our analysis to the cases where the nonlinear magnetoelectric contributions are in one of the two configurations mentioned at the end of Sec.~\ref{Plane-wave}, which are as follows: $(1)$ One assumes that $\beta_{ijk}$ is the only nonzero nonlinear optical coefficient, is diagonal in the direction of the applied magnetic field, and works in the absence of an electric field, and $(2)$ the other assumes that $\gamma_{ijk}$ is the only nonzero nonlinear optical coefficient, is diagonal in the direction of the applied electric field, and works in the absence of a magnetic field. The description of both cases can be encompassed by the effective coefficient $\zeta_{ij}$ that appears in Eq.~(\ref{zetacoe}). For instance, in configuration $(1)$, with a magnetic field in the $a$th direction, this coefficient would read 
\begin{align}
    \zeta_{ij} &= \operatorname{diag}\left(\alpha_{1}+\beta_{11a}B, \alpha_{2}+\beta_{22a}B, \alpha_{3}+\beta_{33a}B\right) \nonumber \\
    & \doteq \operatorname{diag}\left(\zeta_1,\zeta_2,\zeta_3\right).
    \label{zetai}
\end{align}

Now, the Fresnel tensor in Eq.~(\ref{zijnl}) reduces to
\begin{align}
Z_{ij} =\; & \varepsilon_{i j} v^{2} + \left(\epsilon_{ikl} \zeta_{jk} +\epsilon_{jkl} \zeta_{ik} \right) \kappa_{l} v 
\nonumber \\
& -\epsilon_{ikn} \epsilon_{jls} \bar\mu_{kl} \kappa_{n} \kappa_{s},
\label{fres2}
\end{align}
and $\det |Z_{i j}| = 0$ leads to the fourth-degree equation for $v$,
\begin{align}
a_{4} v^{4}+a_{2} v^{2}+a_{1} v + a_{0}=0,
\label{v2}
\end{align}
where
\begin{align}
a_{4}=&\;\varepsilon_{1} \varepsilon_{2} \varepsilon_{3}, 
\nonumber\\
a_{2}=&-\kappa_{1}^{2} \varepsilon_{1}\left(\frac{\varepsilon_{2}}{\mu_{2}}+\frac{\varepsilon_{3}}{\mu_{3}}+\left(\sigma_{23}\right)^{2}\right)
\nonumber \\
&-\kappa_{2}^{2} \varepsilon_{2}\left(\frac{\varepsilon_{1}}{\mu_{1}}+\frac{\varepsilon_{3}}{\mu_{3}}+\left(\sigma_{13}\right)^{2}\right) 
\nonumber \\
&-\kappa_{3}^{2} \varepsilon_{3}\left(\frac{\varepsilon_{1}}{\mu_{1}}+\frac{\varepsilon_{2}}{\mu_{2}}+\left(\sigma_{12}\right)^{2}\right), \nonumber\\
a_{1}=&\; 2 \kappa_{1} \kappa_{2} \kappa_{3}\left(\frac{\varepsilon_1\sigma_{23}}{\mu_{1}}+\frac{\varepsilon_{2}\sigma_{31}}{\mu_{2}}
+\frac{\varepsilon_{3}\sigma_{12}}{\mu_{3}} 
-\sigma_{12}\sigma_{23}\sigma_{31}\right), 
\nonumber\\
a_{0}=&\;\varepsilon_1\left(\frac{\kappa_{1}^{2} \kappa_{2}^{2}}{\mu_{1} \mu_{3}}+\frac{\kappa_{1}^2 \kappa_{3}^{2}}{\mu_{1} \mu_{2}}+\frac{\kappa_{1}^{4}}{\mu_{2} \mu_{3}}\right)
\nonumber \\
&+\varepsilon_{2}\left(\frac{\kappa_{2}^{2} \kappa_{3}^{2}}{\mu_{1} \mu_{2}}+\frac{\kappa_{1}^{2} \kappa_{2}^{2}}{\mu_{2} \mu_{3}}+\frac{\kappa_{2}^{4}}{\mu_{1} \mu_{3}}\right) 
\nonumber \\
 &+\varepsilon_{3}\left(\frac{\kappa_{3}^{2} \kappa_{1}^{2}}{\mu_{3} \mu_{2}}+\frac{\kappa_{3}^2 \kappa_{2}^{2}}{\mu_{3} \mu_{1}}+\frac{\kappa_{3}^{4}}{\mu_{1} \mu_{2}}\right)
\nonumber \\
&+\frac{\kappa_{1}^{2} \kappa_{2}^{2}}{ \mu_{3}}\left(\sigma_{12}\right)^{2}+\frac{\kappa_{2}^{2} \kappa_{3}^{2}}{ \mu_{1}}\left(\sigma_{23}\right)^{2}
 +\frac{\kappa_{1}^{2} \kappa_{3}^{2} }{ \mu_{2}}\left(\sigma_{13}\right)^{2},
 \nonumber
\end{align}
in which we have defined $\sigma_{ij} \doteq \zeta_i - \zeta_j$ and $\hat\kappa_i\hat\kappa_i=1$.
Note that the coefficient $a_1$ survives only when the wave vector has components in the three principal axes. When the propagation is set perpendicularly to one of these axes, or when the magnetoelectric effect is not present, Eq.~(\ref{v2}) will reduce to a biquadratic equation whose solutions will be symmetric under space reversal.  
For instance, if the propagation is set in the $x$ direction ($\kappa_{1}=1$), Eq.~(\ref{v2}) reduces to 
\begin{align}
ab\, v^{4}-(a+b+\Omega)\, v^{2}+1=0,
\label{sd}
\end{align}
where we have defined $a=\varepsilon_{2} \mu_{3},$ $b=\varepsilon_{3} \mu_{2}$, and $\Omega=\mu_{2} \mu_{3} \left(\zeta_{2}-\zeta_{3}\right)^{2}.$

The exact solutions of this algebraic equation can be presented in the form:
\begin{align}
v^{2}_{\scriptscriptstyle \pm} 
= \frac{2}{a+b+\Omega \mp \sqrt{(a+b+\Omega)^{2}-4 a b}} \doteq \frac{c^2}{n_\mp^2},
\label{sds}
\end{align}
where $n_{\pm}$ are the refractive indices experienced by the $\pm$ propagating modes. 
Generally, if $ab >0$, which is the case for natural materials, there will be two distinct solutions for plane-wave propagation in a same direction. In other words, the birefringence effect is present in this type of magnetoelectric material.  

Regarding the birefringence effect, an interesting result emerges from the aforementioned solutions. If we take the product of $v_{\scriptscriptstyle +}$ and $v_{\scriptscriptstyle -}$ in Eq.~(\ref{sds}), it follows that 
\begin{equation}
    v_{\scriptscriptstyle +} v_{\scriptscriptstyle -} = \frac{1}{\sqrt{\varepsilon_2\mu_3}}\frac{1}{\sqrt{\varepsilon_3\mu_2}},
    \label{id}
\end{equation}
which coincides with the product of the birefringence solutions in the absence of the magnetoelectric effect. 
In other words, although each individual phase velocity in the birefringence effect heavily depends on the magnetoelectric coefficient, their product remains independent of it. 
An equivalent form to express the above result would be $v_{\scriptscriptstyle +} v_{\scriptscriptstyle -} = v_{\scriptscriptstyle 2} v_{\scriptscriptstyle 3}$, where $v_\gamma \doteq (\varepsilon_\gamma\mu_\gamma)^{-1/2}$ (here, $\gamma$=2,3, and there is no summation in the repeated index $\gamma$) is effectively the phase velocity of a plane wave in an isotropic dielectric characterized only by the dielectric permittivity $\varepsilon_\gamma$ and the magnetic permeability $\mu_\gamma$.

The phase velocities in Eq.~(\ref{sds}) were obtained for the case of propagation in the $x$ direction. The results for the other two orthogonal directions can be directly found by implementing a cyclic permutation of the indices.  For instance, if the propagation is set in the $y$ direction(i.e., $\kappa_{2}=1$), the parameters in Eq.~(\ref{sds}) should be rewritten by exchanging $\varepsilon_{2} \to \varepsilon_{3}$ and $\varepsilon_{3} \to \varepsilon_{1}$, $ \mu_{2} \to \mu_{3}$ and $\mu_{3} \to \mu_{1}$, and $\zeta_{2} \to \zeta_{3}$ and $\zeta_{3} \to \zeta_{1}$.

\section{Plane waves in a linear medium}
\label{linearmedium}
Now, the analysis is restricted to the realm of linear materials, for which the time-domain constitutive relations in Eq.~(\ref{dh1}) reduce to $D_{i} = \varepsilon_{i j} E_{j} + \alpha_{i j} B_j$ and $H_{i} = \bar\mu_{i j} B_{j} - \alpha_{j i} E_j$. In a frequency-domain description, which is required when a dispersive medium is taken into account, this representation is known \cite{wernerwein,2010eabf.book.....M} as the Boys-Post representation, in contrast to the Tellegen representation used in Appendix \ref{appendix}. 
The study follows a parallel with an earlier work \cite{fuchs1965} in which the fields $\boldsymbol{E}$ and $\boldsymbol{H}$ were assumed to be the variables that determine the polarization and magnetization of the material. 
However, as discussed in Sec.~\ref{FEB}, the variables are here assumed to be $\boldsymbol{E}$ and $\boldsymbol{B}$.

Note that the optical coefficients in the linear sector in Eq.~(\ref{dh1}), which includes $\alpha_{ij}$, do not couple to external fields in the expression for $Z_{ij}$ given by Eq.~(\ref{fres2}). Thus, the results discussed in this section are still valid in the absence of those fields.

Now, the Fresnel tensor reduces to 
$Z_{ij} = \varepsilon_{i j} v^{2} + \left(\epsilon_{ikl} \alpha_{jk} +\epsilon_{jkl} \alpha_{ik} \right)\kappa_l v - \epsilon_{ikn} \epsilon_{jls} \bar\mu_{kl} \kappa_{n} \kappa_{s},$
where the related optical coefficients are given by Eq.~(\ref{coefs}).

Setting the propagation in the $x$ direction, the refractive indices can be immediately obtained from Eq.~(\ref{sds}) and are given by
\begin{align}
\frac{n_\pm^2}{c^2}
= \frac{a+b+\xi \pm \sqrt{(a+b+\xi)^{2}-4 a b}}{2},
\label{refracindices}
\end{align}
where $$\xi = \mu_{2} \mu_{3} \left(\alpha_{2}-\alpha_{3}\right)^{2}$$ and $a$ and $b$ are the same as defined in the last section. 

Unlike the result based on $\boldsymbol{E}$ and $\boldsymbol{H}$ fields \cite{fuchs1965}, here, the magnetoelectric contribution appears only as the square of the difference between $\alpha_{2}$ and $\alpha_{3}$ by means of $\xi$. 

\subsection{Measurable linear magnetoelectric coefficients}
\label{indirect}
Assuming that an experiment is prepared to measure the refractive index of the material, and that its dielectric and magnetic coefficients $\varepsilon_i$ and $\mu_i$ are already known, Eq.~(\ref{refracindices}) can be inverted to obtain $\xi = n^2/c^2+ab c^2/n^2-a-b$. The symbol $\pm$, denoting the two possible solutions in the birefringence effect, was dropped here because $\xi$ does not depend on the specific solution used in its computation, as it can be easily verified. This is expected, as long as $\xi$ represents a measurable quantity that characterizes the optical medium. Thus, the magnetoelectric coefficients are given by 
\begin{align}
(\alpha_{2}-\alpha_{3})^2 =&\; \frac{1}{\mu_{2} \mu_{3}}\left(\frac{\varepsilon_{2} \mu_{3}\varepsilon_{3} \mu_{2}c^2}{n^2}\right.
\nonumber\\
&\left.+\frac{n^2}{c^2}-\varepsilon_{2} \mu_{3}-\varepsilon_{3} \mu_{2}\right).
\nonumber
\end{align}
By choosing the other two orthogonal directions, similar expressions can be obtained for $\alpha_{1}-\alpha_{2}$ and $\alpha_{1}-\alpha_{3}$.

One case of interest is when the material presents optical axes. For instance, in the particular case of $\varepsilon_{ij} = \operatorname{diag}(\varepsilon_{\scalebox{0.6}{\paral}}, \varepsilon_{\scalebox{0.6}{$\perp$}}, \varepsilon_{\scalebox{0.6}{$\perp$}})$ and $\mu_{ij} =\operatorname{diag} (\mu_{\scalebox{0.6}{\paral}}, \mu_{\scalebox{0.6}{$\perp$}}, \mu_{\scalebox{0.6}{$\perp$}})$, it follows that
\begin{align}
|\alpha_{2}-\alpha_{3}| = \frac{1}{\mu_{\scalebox{0.6}{$\perp$}}}\left[\frac{(\varepsilon_{\scalebox{0.6}{$\perp$}} \mu_{\scalebox{0.6}{$\perp$}} c)^2}{n_{x}^2}
+\frac{n_{x}^2}{c^2}-2\varepsilon_{\scalebox{0.6}{$\perp$}} \mu_{\scalebox{0.6}{$\perp$}}\right]^{1/2},
\label{alpha23perp}
\end{align}
where $n_{x}$, in this case, is the refractive index of the medium in the direction of the optical axis. 

A little bit more elaborate expressions can be similarly obtained for the other two perpendicular directions. For propagation in the $y$ direction it follows that
\begin{equation}
\label{alpha13perp}
\begin{split}
 |\alpha_{1}-\alpha_{3}| =&\frac{1}{\mu_{\scalebox{0.6}{$\perp$}}\mu_{\scalebox{0.6}{\paral}}}\left[\varepsilon_{\scalebox{0.6}{\paral}}\mu_{\scalebox{0.6}{\paral}}\varepsilon_{\scalebox{0.6}{$\perp$}}\mu_{\scalebox{0.6}{$\perp$}}\frac{c^2}{n_y^2}\right.\\
 &\left.+\frac{n_y^2}{c^2}-\left(\varepsilon_{\scalebox{0.6}{\paral}}\mu_{\scalebox{0.6}{$\perp$}}+\varepsilon_{\scalebox{0.6}{$\perp$}}\mu_{\scalebox{0.6}{\paral}}\right)\right]^{1/2},
 \end{split}
\end{equation}
while for propagation in the $z$ direction we obtain
\begin{equation}
\label{alpha12perp}
\begin{split}
 |\alpha_{1}-\alpha_{2}|=&\frac{1}{\mu_{\scalebox{0.6}{$\perp$}}\mu_{\scalebox{0.6}{\paral}}}\left[\varepsilon_{\scalebox{0.6}{\paral}}\mu_{\scalebox{0.6}{\paral}}\varepsilon_{\scalebox{0.6}{$\perp$}}\mu_{\scalebox{0.6}{$\perp$}}\frac{c^2}{n_z^2}\right.\\
 &\left.+\frac{n_z^2}{c^2}-\left(\varepsilon_{\scalebox{0.6}{\paral}}\mu_{\scalebox{0.6}{$\perp$}}+\varepsilon_{\scalebox{0.6}{$\perp$}}\mu_{\scalebox{0.6}{\paral}}\right)\right]^{1/2}.
 \end{split}
\end{equation}
Note that if $n_y=n_z$, we find that $\alpha_{2} =\alpha_{3}$ and we can write $\alpha_{ij}=\operatorname{diag}(\alpha_{\scalebox{0.6}{\paral}}, \alpha_{\scalebox{0.6}{$\perp$}}, \alpha_{\scalebox{0.6}{$\perp$}})$; that is, the magnetoelectric effect will present a uniaxial anisotropy, like the permittivity and permeability coefficients. 

An important point here is that, in the context of the linear magnetoelectric effect, the results expressed by Eqs.~(\ref{alpha23perp}), (\ref{alpha13perp}) and (\ref{alpha12perp}) are exact. No approximations were required to derive these equations. 

\subsection{Regime with a weak magnetoelectric contribution}
There are two situations that must be mentioned when a weak magnetoelectric regime is assumed.
First, when $a \neq b$, the phase velocities will depend on the square of the magnetoelectric coefficient. 
In fact, if $\xi$ is small compared to $a$ or $b$, it follows from Eq.~(\ref{refracindices}) that
\begin{subequations}
\label{vb+-}
\begin{align}
v_{\scriptscriptstyle +}^2&\simeq \frac{1}{\varepsilon_{2} \mu_{3}}\left[1+\frac{\mu_{2} \mu_{3}\left(\alpha_{2}-\alpha_{3}\right)^{2}}{\left(\varepsilon_{2} \mu_{3}-\varepsilon_{3} \mu_{2}\right)}\right],
\label{vb+}
\\
v_{\scriptscriptstyle -}^2 & \simeq  \frac{1}{\varepsilon_{3} \mu_{2}}\left[1-\frac{\mu_{2} \mu_{3}\left(\alpha_{2}-\alpha_{3}\right)^{2}}{\left(\varepsilon_2 \mu_{3}-\varepsilon_{3} \mu_{2}\right)}\right],
\label{vb-}
\end{align}
\end{subequations}
which are quadratic in the difference $\alpha_{2}-\alpha_{3}$. 

It is interesting to note that the magnetoelectric effect makes a contribution to the already present birefringence effect due to the purely dielectric sector if $\varepsilon_2 \neq \varepsilon_3$ or the magnetic sector if $\mu_{2}\neq\mu_{3}$.

On the other hand, if $a = b$ birefringence will occur only when the magnetoelectric contribution is present, as can be inferred from Eq.~(\ref{refracindices}). Furthermore, the effect will be proportional to $\alpha_{2}-\alpha_{3}$,  
\begin{equation}
v_{\scriptscriptstyle \pm}^2\simeq \frac{1}{a}\left(1\pm\sqrt{\frac{\xi}{a}}\right);
\nonumber
\end{equation}
i.e., it will depend on the first-order power of the magnetoelectric coefficient. This is particularly the case, for instance, when the system is isotropic in the electric and magnetic sectors, $\chi_{i j}=\chi\delta_{i j}$ and $\tilde\chi_{i j}=\tilde\chi\delta_{i j}$, but with $\alpha_{2}\neq \alpha_{3}$, which leads to 
\begin{align}
\label{vbapprox}
v_{\scriptscriptstyle \pm}^2 \simeq \frac{1}{\varepsilon\mu}\left(1
\pm\frac{\mu}{\sqrt{\varepsilon\mu}}\left|\alpha_{2}-\alpha_{3}\right| \right).
\end{align}

Similar results can easily be obtained for propagation in the other two orthogonal directions.
More complicated arrangements, as in the case of propagation in an arbitrary direction, require solving Eq.~(\ref{v2}) with $a_1 \neq 0$. Furthermore, if the principal axes of the three optical coefficients do not coincide, the general eigenvalue problem must be considered. 

\section{Final remarks}
\label{final}
It is instructive to note how the magnetoelectric contributions manifest in the $Z_{ij}$ tensor, impacting the algebraic equation for phase velocity. By examining Eq.~(\ref{thzela}), we can observe the distinct role of each optical coefficient in the behavior of light propagation. First, the term proportional to $v$ in the Fresnel tensor is related only to magnetoelectric contributions, both linear and nonlinear, with the linear contribution appearing only in this term, while the nonlinear contributions also manifest in the other two terms of this tensor. When the determinant of $Z_{ij}$ is calculated, this term will result in odd-power terms in the algebraic equation. In the particular case of a linear medium, examined in Sec.~\ref{linearmedium}, it becomes clear that such terms are responsible for a break in the symmetry of light propagating in opposite directions. Second, as previously stated, the nonlinear magnetoelectric coefficients also appear in other terms, namely, $\theta_{ij}$ and $\lambda_{ij}$. These contributions can be seen separately because they impact different optical sectors of the system. While the magnetic field induces a correction to the electric permittivity by means of its coupling to $\gamma_{ijk}$, the electric field leads to modifications of the magnetic permeability by means of its coupling to $\beta_{ijk}$. Both of these effects contribute to the birefringence in light propagation.

Regarding the possible impacts of the results in the experimental context, there are some remarks worth making. An important observable quantity in optics is the refractive index $n$ of the medium. The optical coefficients can be obtained by measuring $n$ for specific directions of light propagation. By studying particular setups of light propagation in nonlinear magnetoelectrics, based on the description presented here, one can find relations that enable the measurement of all components of the coefficients \(\beta_{ijk}\) and \(\gamma_{ijk}\), which expand upon previous analyses conducted with certain approximations \cite{PhysRevA.105.023530}. Additionally, we should observe that the values of these coefficients in the description adopted here will certainly be quite different from those obtained when the description is based on \(\boldsymbol{E}\) and \(\boldsymbol{H}\) fields. As discussed, even their physical dimensions are distinct in both descriptions. In order to compare both coefficients, we should first relate them by comparing the corresponding constitutive relations. These comparisons will lead to a generalization of the Boys-Post and Tellegen descriptions by also including the nonlinear coefficients and deserve further investigation.

When only the linear case is considered, the magnetoelectric contributions arise solely through the coefficient $\xi$, which is a quadratic function of $\alpha_{i}$. In particular, the results in Eqs.~(\ref{vb+-}) and (\ref{vbapprox}) were obtained by assuming small $\xi$, which means that the difference between the two components of the magnetoelectric coefficient is small compared to the other linear coefficients. However, the same results hold for the case of small $\alpha_i$, which is a less restrictive assumption. On the other hand, when an analysis based on the traditional description $F(\boldsymbol{E},\boldsymbol{H};T)$ is done, as presented in Appendix \ref{appendix}, the results are a little bit more involved, as the magnetoelectric coefficients appear in two different functions ($\epsilon^{\scriptscriptstyle \boldsymbol{E\!H}}_i$ and $\xi'$) with different functional dependence \cite{fuchs1965}. In that case, approximations assuming small $\xi'$  or small $\alpha^{\scriptscriptstyle \boldsymbol{E\!H}}_i$ could correspond to different physical regimes.

Finally, as discussed in Appendix~\ref{magsus}, defining the magnetic susceptibility in terms of the coefficient of $\boldsymbol{B}$ or $\boldsymbol{H}$ in the expansion of the magnetization vector leads to alternative ways of expressing some well-known results of electromagnetism. Here, it is suggested that the definition of this quantity based on the prescription in which $\boldsymbol{B}$ is the variable of the free-energy density seems to be a more convenient choice. Additionally, although the magnetic permeabilities in both prescriptions coincide, that is not the case for the other two sectors of the constitutive relations \cite{wernerwein}. In fact, comparing Eqs.~(\ref{dh1}) and (\ref{dh2}), we can see that the prescription based on $\boldsymbol{H}$ leads to an effective electric permittivity $\epsilon_{ij}^{\scriptscriptstyle \boldsymbol{E\!H}}$ \cite{fuchs1965} that mixes the electric, magnetic, and magnetoelectric (up to second order) coefficients and to a magnetoelectric sector that also depends on the magnetic permeability of the material. The relationship between the magnetoelectric coefficients of both approaches can be obtained directly from these equations, resulting in $\alpha_{ij} =  \bar\mu^{\scriptscriptstyle \boldsymbol{E\!H}}_{j k}\alpha^{\scriptscriptstyle \boldsymbol{E\!H}}_{i k}$. Particularly, for the case of a diagonal system, the relations $\mu_0\alpha_i=\alpha^{\scriptscriptstyle \boldsymbol{E\!H}}_i/(1+\tilde\chi^{\scriptscriptstyle \boldsymbol{E\!H}}_i)$ and $\alpha^{\scriptscriptstyle \boldsymbol{E\!H}}_i=\mu_0\alpha_i/(1-\tilde\chi_i)$ follow. The electric sectors in both prescriptions coincide when only first-order effects are considered.

\begin{acknowledgments}
V.~A.~D.~L. acknowledges partial support from the Conselho Nacional de Desenvolvimento Científico e Tecnológico (CNPq), Brazil, Grant No. 302492/2022-4. L.~T.~P. is funded by Fundação de Amparo à Pesquisa do Estado de São Paulo (FAPESP), Brazil, Grant No. 23/07013-2.
\end{acknowledgments}

\section*{Data Availability}
No datasets were generated or analysed during the current study.

\appendix

\section{Description with \texorpdfstring{$F(\boldsymbol{E}, \boldsymbol{H}; T)$}{F(E, H; T)} for the linear case}

\label{appendix}
The choice of $\boldsymbol{E}$ and $\boldsymbol{H}$ as thermodynamic variables for $F$ is the one traditionally used in the literature. 
In this case, ignoring spontaneous processes, which do not contribute to the effects examined here, the expansion of $F$ in terms of these fields is given by \cite{1973-schmid,rivera1994a,2009EPJB...71..299R,2005JPhD...38R.123F}
\begin{align}
F\left(\boldsymbol{E}, \boldsymbol{H};T\right) =&\,
F_{0}-\tfrac{1}{2}\varepsilon_0\chi_{i j}^{\scriptscriptstyle \boldsymbol{E\!H}} E_{i} E_{j} -\tfrac{1}{2}\mu_{0} \tilde\chi_{i j}^{\scriptscriptstyle \boldsymbol{E\!H}} H_{i} H_{j}
\nonumber\\
&-\alpha_{i j}^{\scriptscriptstyle \boldsymbol{E\!H}} E_{i} H_{j},
\label{freeEH}
\nonumber
\end{align}
where the electric and magnetic susceptibilities of the medium are now represented, respectively, by the dimensionless coefficients $\chi_{i j}^{\scriptscriptstyle \boldsymbol{E\!H}}$ and $\tilde\chi_{i j}^{\scriptscriptstyle \boldsymbol{E\!H}}$, and the magnetoelectric coefficient in this description is given by $\alpha_{i j}^{\scriptscriptstyle \boldsymbol{E\!H}}$, whose physical dimension is the inverse of velocity. Now the superscript {${\boldsymbol{E\!H}}$} was added to make a link to the use of the representation $F(\boldsymbol{E},\boldsymbol{H};T)$.

The polarization and magnetization vectors can now be obtained by means of the derivatives of $F$ as follows: $P_{i}=-(\partial F/\partial E_{i})$, and $\mu_{0} M_{i}=-(\partial F/\partial H_{i})$, which leads to constitutive relations in the Tellegen representation \cite{wernerwein,2010eabf.book.....M}, $D_{i} = \varepsilon_{i j}^{\scriptscriptstyle \boldsymbol{E\!H}} E_{j}+\alpha^{\scriptscriptstyle \boldsymbol{E\!H}}_{i j} H_j$ and $B_{i} = \mu^{\scriptscriptstyle \boldsymbol{E\!H}}_{i j} H_{j}+\alpha^{\scriptscriptstyle \boldsymbol{E\!H}}_{j i} E_{j}$,
where we defined $\varepsilon_{i j}^{\scriptscriptstyle \boldsymbol{E\!H}} \doteq \varepsilon_0\left(\delta_{ij} + \chi_{i j}^{\scriptscriptstyle \boldsymbol{E\!H}}\right)$, and $\mu^{\scriptscriptstyle \boldsymbol{E\!H}}_{i j}$  is the magnetic permeability tensor defined by Eq.~(\ref{muH}). Now, the constitutive relations can be conveniently written as follows:
\begin{subequations}
\label{dh2}
\begin{align}
D_{i} &= \epsilon_{i j}^{\scriptscriptstyle \boldsymbol{E\!H}} E_{j} + \bar\mu^{\scriptscriptstyle \boldsymbol{E\!H}}_{k j}\alpha^{\scriptscriptstyle \boldsymbol{E\!H}}_{i k} B_j,
\\
H_{i}&=\bar\mu^{\scriptscriptstyle \boldsymbol{E\!H}}_{i j} B_{j}-\bar\mu^{\scriptscriptstyle \boldsymbol{E\!H}}_{i k} \alpha^{\scriptscriptstyle \boldsymbol{E\!H}}_{j k} E_{j},
\end{align}
\end{subequations}
where the electric permittivity tensor is identified as 
\begin{align}
\epsilon_{i j}^{\scriptscriptstyle \boldsymbol{E\!H}} \doteq \varepsilon_{i j}^{\scriptscriptstyle \boldsymbol{E\!H}} - \alpha^{\scriptscriptstyle \boldsymbol{E\!H}}_{il} \bar\mu^{\scriptscriptstyle \boldsymbol{E\!H}}_{l k}  \alpha^{\scriptscriptstyle \boldsymbol{E\!H}}_{jk}.
\nonumber
\end{align}

If we compare these relations with the equivalent ones obtained in Sec.~\ref{FEB}, we see that here, a coupling appears between the magnetic and magnetoelectric coefficients.

Now, if we repeat the steps detailed in Sec.~\ref{linearmedium}, we obtain the general eigenvalue problem $Z^{\scriptscriptstyle \boldsymbol{E\!H}}_{ij}e_j = 0,$ where
\begin{align}
Z^{\scriptscriptstyle \boldsymbol{E\!H}}_{ij} =\; & \epsilon_{i j}^{\scriptscriptstyle \boldsymbol{E\!H}} v^{2} +\bar\mu^{\scriptscriptstyle \boldsymbol{E\!H}}_{kl} \left(\epsilon_{k n i} \alpha^{\scriptscriptstyle \boldsymbol{E\!H}}_{jl} +\epsilon_{k n j} \alpha^{\scriptscriptstyle \boldsymbol{E\!H}}_{il} \right) \kappa_{n} v 
\nonumber \\
& -\epsilon_{ikn} \epsilon_{jls} \bar\mu^{\scriptscriptstyle \boldsymbol{E\!H}}_{kl} \kappa_{n} \kappa_{s}.
\nonumber
\end{align}

The analysis is now restricted to systems whose optical coefficients are diagonal tensors, which means that their principal axes can be chosen to be coincident, namely,
\begin{align}
    \mu^{\scriptscriptstyle \boldsymbol{E\!H}}_{i j}&=\operatorname{diag}\left({\mu^{\scriptscriptstyle \boldsymbol{E\!H}}_{1}}, {\mu^{\scriptscriptstyle \boldsymbol{E\!H}}_{2}},{\mu^{\scriptscriptstyle \boldsymbol{E\!H}}_{3}}\right),
    \nonumber
    \\
   \alpha^{\scriptscriptstyle \boldsymbol{E\!H}}_{i j}&=\operatorname{diag}\left(\alpha^{\scriptscriptstyle \boldsymbol{E\!H}}_{1}, \alpha^{\scriptscriptstyle \boldsymbol{E\!H}}_{2}, \alpha^{\scriptscriptstyle \boldsymbol{E\!H}}_{3}\right),
    \nonumber
    \\
    \epsilon_{i j}^{\scriptscriptstyle \boldsymbol{E\!H}}&=\operatorname{diag}\left(\epsilon_{1}^{\scriptscriptstyle \boldsymbol{E\!H}}, \epsilon_{2}^{\scriptscriptstyle \boldsymbol{E\!H}}, \epsilon_{3}^{\scriptscriptstyle \boldsymbol{E\!H}}\right),
    \nonumber
\end{align}
where $\epsilon_{i}^{\scriptscriptstyle \boldsymbol{E\!H}} = \varepsilon_{i}^{\scriptscriptstyle \boldsymbol{E\!H}}-{\left(\alpha^{\scriptscriptstyle \boldsymbol{E\!H}}_i\right)^2}/{\mu^{\scriptscriptstyle \boldsymbol{E\!H}}_i}$, for $i=1,2,3$.

Then, setting the particular propagation direction $\kappa_{x}=1$ and solving $\det |Z_{ij}|=0$ for $v$ result \cite{fuchs1965} in a biquadratic equation whose solutions for the refractive indices are given as in Eq.~(\ref{refracindices}), but with $a\rightarrow  a'$, $b\rightarrow b'$, and $\xi \rightarrow \xi'$, where
\begin{align}
  a' &= \epsilon_{2}^{\scriptscriptstyle \boldsymbol{E\!H}} \mu_{3}^{\scriptscriptstyle \boldsymbol{E\!H}} = \left[\varepsilon_{2}^{\scriptscriptstyle \boldsymbol{E\!H}}-\frac{\left(\alpha^{\scriptscriptstyle \boldsymbol{E\!H}}_2\right)^2}{\mu^{\scriptscriptstyle \boldsymbol{E\!H}}_2} \right]\mu_{3}^{\scriptscriptstyle \boldsymbol{E\!H}}, \nonumber \\
b' &=\epsilon_{3}^{\scriptscriptstyle \boldsymbol{E\!H}} \mu_{2}^{\scriptscriptstyle \boldsymbol{E\!H}} = \left[\varepsilon_{3}^{\scriptscriptstyle \boldsymbol{E\!H}}-\frac{\left(\alpha^{\scriptscriptstyle \boldsymbol{E\!H}}_3\right)^2}{\mu^{\scriptscriptstyle \boldsymbol{E\!H}}_3}\right] \mu_{2}^{\scriptscriptstyle \boldsymbol{E\!H}}, \nonumber\\
\xi' &=\mu^{\scriptscriptstyle \boldsymbol{E\!H}}_2\mu^{\scriptscriptstyle \boldsymbol{E\!H}}_3\left(\frac{\alpha^{\scriptscriptstyle \boldsymbol{E\!H}}_2}{\mu^{\scriptscriptstyle \boldsymbol{E\!H}}_2} -\frac{\alpha^{\scriptscriptstyle \boldsymbol{E\!H}}_3}{\mu^{\scriptscriptstyle \boldsymbol{E\!H}}_3}\right)^{2}.
\nonumber
\end{align}
Note that unlike $a$ and $b$ in Eq.~(\ref{refracindices}), $a'$ and $b'$ are also functions of the magnetoelectric coefficient.

\section{On the definition of the magnetic susceptibility}
\label{magsus}
Historically, the magnetic susceptibility has been defined as the proportionality coefficient between the magnetization of the medium and the auxiliary field $\boldsymbol{H}$. However, it is well understood that, fundamentally, the magnetization phenomenon is a consequence of the presence of an applied magnetic field  $\boldsymbol{B}$, and adopting a functional relation such as $\boldsymbol{M}=\boldsymbol{M}(\boldsymbol{B})$ seems to be a natural choice. A comparison between this approach and the traditional one based on $\boldsymbol{M}=\boldsymbol{M}(\boldsymbol{H})$ is discussed below. 

\subsection{Magnetic susceptibility defined through \texorpdfstring{$M(B)$}{M(B)}}
As discussed in Sec.~{\ref{FEB}}, when a nonconducting crystalline material is under the action of an external magnetic field $\boldsymbol{B}$, it will store energy due to the action of that field, and the induced magnetization will be described by the second term on the right-hand side of Eq.~(\ref{&m}).
Thus, using that result, we obtain the constitutive relation  
\begin{align}
B_{i}=\mu_{i j} H_{j},
\label{bmh}
\end{align}
where the magnetic permeability tensor $\mu_{i j}$ is the inverse of the tensor defined in Eq.~(\ref{varmu}).
This constitutive relation can be presented in its classical form as $H_{i}=\bar\mu_{i j} B_{j}$.

As $\tilde\chi_{i j} = \tilde\chi_{j i}$, if we conveniently choose coordinate axes that coincide with the principal axes of the system, this tensor can be written in its diagonal form. In this case $\tilde\chi_{i j}= \operatorname{diag}\left(\tilde\chi_1,\tilde\chi_2,\tilde\chi_3\right)$, where $\tilde\chi_i$ ($i=1,2,3$) are the principal values of the magnetic susceptibility. Then, the magnetic permeability tensor reduces to 
\begin{align}
    \frac{\mu_{i j}}{\mu_0}&=\operatorname{diag}\left(\frac{1}{1 -\tilde\chi_{1}}, \frac{1}{1 -\tilde\chi_{2}},\frac{1}{1 -\tilde\chi_{3}}\right).
    \nonumber
\end{align}
Like the susceptibility tensor, $\mu_{i j}$ is diagonal and has principal values given by the three diagonal components above.

\subsection{Magnetic susceptibility defined through \texorpdfstring{$M(H)$}{M(H)}}
Now we move to the traditional description in which the auxiliary field $\boldsymbol{H}$ is used as a thermodynamic variable for the free-energy density, as discussed in Appendix \ref{appendix}. As a consequence, the magnetization vector will be given by $M_{i}= \tilde\chi_{i j}^{\scriptscriptstyle \boldsymbol{E\!H}} H_{j}$,
and the constitutive relation reduces to
\begin{equation}
B_{i} = \mu^{\scriptscriptstyle \boldsymbol{E\!H}}_{i j} H_{j},
\label{bmh2}
\end{equation}
where the magnetic permeability tensor is defined as 
\begin{equation}
\mu^{\scriptscriptstyle \boldsymbol{E\!H}}_{i j}=\mu_{0}\left(\delta_{i j}+\tilde\chi^{\scriptscriptstyle \boldsymbol{E\!H}}_{i j}\right).
\label{muH}
\end{equation}
For completeness, we introduce the inverse-permeability tensor $\bar\mu^{\scriptscriptstyle \boldsymbol{E\!H}}_{j k}$, so that Eq.~(\ref{bmh2}) can be presented in the canonical form  
$H_{i}=\bar\mu^{\scriptscriptstyle \boldsymbol{E\!H}}_{i j} B_{j}$. 

As before, using the fact that $\tilde\chi_{i j}^{\scriptscriptstyle \boldsymbol{E\!H}}$ is a symmetric rank-2 tensor and choosing a coordinate system in which it is diagonal, we define $\tilde\chi^{\scriptscriptstyle \boldsymbol{E\!H}}_{i j}= \operatorname{diag}\left(\tilde\chi^{\scriptscriptstyle \boldsymbol{E\!H}}_1,\tilde\chi^{\scriptscriptstyle \boldsymbol{E\!H}}_2,\tilde\chi^{\scriptscriptstyle \boldsymbol{E\!H}}_3\right)$.
Then, the magnetic permeability tensor reduces to 
\begin{align}
    \frac{\mu^{\scriptscriptstyle \boldsymbol{E\!H}}_{i j}}{\mu_0}&=\operatorname{diag}\left(1 +\tilde\chi^{\scriptscriptstyle \boldsymbol{E\!H}}_{1},1 +\tilde\chi^{\scriptscriptstyle \boldsymbol{E\!H}}_{2},1 +\tilde\chi^{\scriptscriptstyle \boldsymbol{E\!H}}_{3}\right).
    \nonumber
\end{align}

\subsection{Comparing the prescriptions}
It can be inferred from Eqs.~(\ref{bmh}) and (\ref{bmh2}) that the magnetic permeability from both prescriptions must be equated, that is, $\mu^{\scriptscriptstyle \boldsymbol{E\!H}}_{i j}=\mu_{i j}$, leading to an expression relating the corresponding susceptibilities, 
\begin{align}
    \delta_{ij} + \tilde\chi^{\scriptscriptstyle \boldsymbol{E\!H}}_{ij} = (\delta_{ij} -\tilde\chi_{ij})^{-1}.
    \nonumber
\end{align}
Finally, as the above susceptibilities can be represented as diagonal tensors, it follows that
\begin{align}
    \tilde\chi_{i} = \frac{\tilde\chi^{\scriptscriptstyle \boldsymbol{E\!H}}_{i}}{1+\tilde\chi^{\scriptscriptstyle \boldsymbol{E\!H}}_{i}}.
    \nonumber
\end{align}
For an isotropic medium the subscript $i$ can be omitted.

As an application, the well-known result for the magnetization current density in an isotropic medium, characterized by the magnetic susceptibility $\tilde\chi^{\scriptscriptstyle \boldsymbol{E\!H}}$, under the effect of an applied electric field is given by 
\begin{align}
   \mu_0 J_i(\vec r, t) = \frac{\tilde\chi^{\scriptscriptstyle \boldsymbol{E\!H}}}{1+\tilde\chi^{\scriptscriptstyle \boldsymbol{E\!H}}}\int_{-\infty}^{t} \epsilon_{iln}\epsilon_{jkn}\partial_j\partial_l E_k(\vec r,t') dt'.
    \nonumber
\end{align}
Note that the factor appearing on the right-hand side of this result is not $\tilde\chi^{\scriptscriptstyle \boldsymbol{E\!H}}$, but $\tilde\chi^{\scriptscriptstyle \boldsymbol{E\!H}}/(1+\tilde\chi^{\scriptscriptstyle \boldsymbol{E\!H}}) = \tilde\chi$. This formula suggests that the definition of the magnetic susceptibility $\tilde\chi$, as discussed in Sec.~\ref{FEB}, seems to be more natural. 

We can find a similar comparison between both approaches when we consider the expression for the magnetic energy density of a permeable body placed in a region of constant magnetic field $\boldsymbol{B}$. This quantity can be expressed as $u_M=\tfrac{1}{2}\boldsymbol{M}\cdot\boldsymbol{B}$, and in the case of an isotropic medium, it reduces to
\begin{equation}
    u_M=\frac{\tilde\chi^{\scriptscriptstyle \boldsymbol{E\!H}}}{1+\tilde\chi^{\scriptscriptstyle \boldsymbol{E\!H}}}\frac{{B}^2}{2\mu_0}=\frac{\tilde\chi{B}^2}{2\mu_0}.
    \nonumber
\end{equation}
Once again, the result is a linear function of the magnetic susceptibility $\tilde\chi$.

\section{Preliminary remarks on the optical indicatrix for magnetoelectric media }
\label{ellipsoid}
The index ellipsoid, also known as the optical indicatrix or dielectric ellipsoid, is a fundamental concept in crystal optics. It provides a geometric representation of how light propagates through anisotropic media, particularly in birefringent crystals \cite{1999poet.book.....B, landau1984electrodynamics}, provided some symmetry conditions are verified. 

Previous studies investigated the index ellipsoid in a range of different media, including reciprocal materials \cite{McCall_2016} and dielectrics exhibiting nonlinear effects \cite{PhysRevA.61.053808}. However, the role of magnetoelectric effects in shaping the index ellipsoid remains unexplored. Here, we extend the analysis of light propagation in anisotropic media~\cite{landau1984electrodynamics} to include nonlinear magnetoelectric coefficients in the index ellipsoid analysis. For the sake of simplicity, hereafter, we set $c=1$.

For a plane wave propagating in a material medium, Maxwell's equations lead to $\omega B_i=\epsilon_{ijk}q_jE_k$ and $\omega D_i =-\epsilon_{ijk}q_jH_k$.
Then, defining $\boldsymbol{n}$ through $\boldsymbol{q}=\omega\boldsymbol{n}$, we obtain  \cite{landau1984electrodynamics} 
\begin{subequations}
\label{DBn}
    \begin{align}
        B_i &=\epsilon_{ijk}n_jE_k, \label{Bn}\\
          D_i&=-\epsilon_{ijk}n_jH_k.\label{Dn}
    \end{align}
\end{subequations}
Let us substitute Eq.~(\ref{Bn}) in (\ref{h1}) and write
\begin{equation}
    H_i=(\mathcal{M}_{in}\epsilon_{nlj}n_l+\tilde{\mathcal{M}}_{ij})E_j,
    \label{herelation}
\end{equation}
where
\begin{subequations}
\label{mcoefficients}
\begin{align}
\mathcal{M}_{ij}&=\bar\mu_{i j}-\tfrac{1}{2}\tilde\chi_{ijk}B_k,
\label{C3a}\\
\tilde{\mathcal{M}}_{ij}&=-\left(\alpha_{j i}+\beta_{jik}B_k+\tfrac{1}{2}\gamma_{ijk}E_k\right).
\end{align}
\end{subequations}
Note that the tensor $\mathcal{M}_{ij}$ presents only magnetic coefficients, while $\tilde{\mathcal{M}}_{ij}$ is purely magnetoelectric. Using the relation given by Eq.~(\ref{herelation}) in (\ref{Dn}), we find the relation between the displacement vector and the refractive index, namely,
\begin{align}
D_i =& - \epsilon_{ijk} n_j \tilde{\mathcal{M}}_{kl} E_l - n_i\left( n_l \mathcal{M}_{kk} - n_n \mathcal{M}_{ln} \right)  E_l
\nonumber\\
&+ \left(n^2\delta_{kl} - n_kn_l\right)\left(\mathcal{M}_{kl}E_i-\mathcal{M}_{ki}E_l\right) . 
\label{Dgeneralized}
\end{align}
Note that for the case of nonmagnetic and nonmagnetoelectric materials, i.e.,  $\mathcal{M}_{ij}=\tfrac{\delta_{ij}}
{\mu_0}$ and $\tilde{\mathcal{M}}_{ij}=0$, respectively, we obtain $D_i =\tfrac{1}{\mu_0}\left(n^2\delta_{ij}-n_in_j\right)E_j$.

In what follows, we focus on situations where the nonlinear sector of Eq.~(\ref{p&m}) is influenced solely by magnetoelectric effects, described by $\beta_{ijk}$ and $\gamma_{ijk}$. In addition, we set $\mathcal{M}_{ij}=\delta_{ij}/\mu$.  In this case, Eq.~(\ref{Dgeneralized}) reads
\begin{equation}
    \boldsymbol{D}=\tfrac{1}{\mu}\left[n^2\boldsymbol{E}-(\boldsymbol{n}\cdot\boldsymbol{E})\boldsymbol{n}\right]-\boldsymbol{n}\times\left(\tilde{\mathcal{{M}}}\boldsymbol{E}\right),
\end{equation}
where $\tilde{\mathcal{M}}$ is a matrix whose elements are given by $\tilde{\mathcal{M}}_{ij}$ and $(\tilde{\mathcal{M}}\boldsymbol{E})_i=\mathcal{\tilde{M}}_{ij}E_j$. It is worth noting that the magnetoelectric effect leads to a new contribution to the transverse component of the displacement vector with respect to the direction of $\boldsymbol{n}$. Indeed, we can write the transverse displacement field as
\begin{equation}
    \boldsymbol{D}_\perp=\tfrac{1}{\mu}n^2\boldsymbol{E}_\perp-\boldsymbol{n}\times\Big(\tilde{\mathcal{{M}}}\boldsymbol{E}\Big)_{\scalebox{0.6}{$\perp$}}.
    \label{transversecompon}
\end{equation}

Now, using Eq.~(\ref{Bn}) in (\ref{d1}) we obtain
\begin{equation}
D_i=\Pi_{ij}E_j, 
\label{relationDE}
\end{equation}
where  we defined
\begin{equation}
\Pi_{ij}=\varepsilon_{ij}+\tilde{\varepsilon}_{il}\epsilon_{lnj}n_n,
\end{equation}
with 
$\tilde{\varepsilon}_{il}=\alpha_{il}+\tfrac{1}{2}\beta_{ilk}B_{k}+\gamma_{lik}E_k$.

Let us now invert the relation in Eq.~(\ref{relationDE}) and write it in matrix notation as $\boldsymbol{E}=\Pi^{-1}\boldsymbol{D}$, such that the transverse component (\ref{transversecompon}) can be rearranged as follows:
\begin{equation}
    \Bigg[\frac{1}{n^2}\left(\mathbb{I}~\cdot\right)_{\scalebox{0.6}{$\perp$}}-\frac{1}{\mu}\left(\Pi^{-1}~\cdot\right)_{\scalebox{0.6}{$\perp$}}+\frac{1}{n^2}\boldsymbol{n}\times\left(\tilde{\mathcal{M}}\Pi^{-1}~\cdot\right)_{\scalebox{0.6}{$\perp$}}\Bigg]\boldsymbol{D}=0,
    \label{ellipsoideq}
\end{equation}
where $\mathbb{I}$ is the $(3\times3)$ identity matrix and $(~~\cdot)_{\scalebox{0.6}{$\perp$}}$ denotes an operation applied to the displacement field $\boldsymbol{D}$ which takes its transverse component with respect to $\boldsymbol{n}$. 

In the nonmagnetic and non-magnetoelectric case, i.e., $\Pi_{ij}=\varepsilon_{ij}$,  $\mathcal{\tilde{M}}_{ij}=0$, and $\mu=\mu_0$, we recover the well-known result \cite{landau1984electrodynamics}
\begin{equation}
    \left[\frac{1}{n^2}(\delta_{ij}~~\cdot)_\perp-\frac{1}{\mu_0}(\varepsilon_{ij}^{-1}~~\cdot)_\perp\right]D_i=0.
    \label{}
\end{equation}
In this particular scenario, since the tensor $\varepsilon^{-1}_{ij}$ is symmetric, it admits a set of orthogonal principal axes, along which $\boldsymbol{D}$ is naturally projected. This allows us to construct an ellipsoid associated with this tensor, providing a geometric interpretation of light propagation in the material.

When more general materials are considered, the resulting surface associated with the optical tensor in (\ref{ellipsoideq}) might correspond to a general quadric. The departure from the standard index-ellipsoid description can be anticipated from the fact that the effective tensor governing the optical response may be nonsymmetric in the presence of magnetoelectric coupling. In such cases, the existence of a complete set of orthogonal principal directions is not ensured, and the polarization modes may not exhibit the usual orthogonal projection onto well-defined axes. As a consequence, the conventional geometric interpretation of the index surface in terms of principal axes is not straightforward. A thorough analysis of this generalized scenario lies beyond the scope of the present work and is worth further investigation. 

\bibliography{optics}

\end{document}